\documentclass[conference]{IEEEtran}
\IEEEoverridecommandlockouts
\usepackage{cite}
\usepackage{url}
\usepackage{amsmath,amssymb,amsfonts}
\usepackage{algorithmic}
\usepackage{graphicx}
\usepackage{textcomp}
\usepackage{xcolor}
\usepackage{subfig}
\def\BibTeX{{\rm B\kern-.05em{\sc i\kern-.025em b}\kern-.08em
    T\kern-.1667em\lower.7ex\hbox{E}\kern-.125emX}}
\begin{document}

\title{A Multi-Stage Framework for Multimodal Controllable Speech Synthesis\thanks{This work is supported by National Natural Science Foundation of China (62076144) and Shenzhen Science and Technology Program (JCYJ20220818101014030).}}

\author{
\IEEEauthorblockN{
Rui Niu\textsuperscript{1},
Weihao Wu\textsuperscript{1},
Jie Chen\textsuperscript{2},
Long Ma\textsuperscript{2},
Zhiyong Wu\textsuperscript{1}\IEEEauthorrefmark{1}\thanks{\IEEEauthorrefmark{1}Corresponding author.}
}
\IEEEauthorblockA{\textsuperscript{1}Shenzhen International Graduate School, Tsinghua University, Shenzhen, China\\
}

\IEEEauthorblockA{\textsuperscript{2}Youtu Lab, Tencent, Beijing, China\\
\{nr23, wuwh23\}@mails.tsinghua.edu.cn, zywu@sz.tsinghua.edu.cn}
godjchen@tencent.com, malong@gmail.com
}

\maketitle

\begin{abstract}
Controllable speech synthesis aims to control the style of generated speech using reference input, which can be of various modalities.
Existing face-based methods struggle with robustness and generalization due to data quality constraints, while text prompt methods offer limited diversity and fine-grained control. 
Although multimodal approaches aim to integrate various modalities, their reliance on fully matched training data significantly constrains their performance and applicability.
This paper proposes a 3-stage multimodal controllable speech synthesis framework to address these challenges. 
For face encoder, we use supervised learning and knowledge distillation to tackle generalization issues. 
Furthermore, the text encoder is trained on both text-face and text-speech data to enhance the diversity of the generated speech. 
Experimental results demonstrate that this method outperforms single-modal baseline methods in both face based and text prompt based speech synthesis, highlighting its effectiveness in generating high-quality speech.
\end{abstract}


\section{Introduction}
In recent years, remarkable progress has been made in text-to-speech (TTS) synthesis technology. 
The intelligibility and naturalness of the generated speech are now nearly indistinguishable from those of humans.
Meanwhile, how to control the style of generated speech to meet personalized needs poses a challenge to existing speech synthesis systems. 
% The synthesis of speech for unseen speakers has received increasing attention.
Specifically, controllable speech synthesis aims to extract timbre, prosody, and style information from reference input, and generate highly consistent and realistic speech from text based on these information. 

Depending on the modality of reference input, controllable speech synthesis can be categorized into four types: reference speech based, facial image based, text prompt based and multimodal based.
Methods based on reference speech\cite{jiang2024mega,du2024cosyvoice,wang2024maskgct} extract timbre and prosody information from a few seconds speech of target speaker.
The advancement of large-scale speech data and model architectures has resulted in a notable improvement in the performance of these methods.
In face based speech synthesis\cite{10094745,10448433,goto2020face2speech,lee24_interspeech}, speaker-related information derived from a facial image is employed. 
Due to data quality constraints, these methods encounter challenges in terms of robustness and generalization ability.
% chenjie: Text.prompt based methods, 应与上文一致，不然会混淆
Text prompt based methods\cite{yang2024instructtts,guo2023prompttts,shimizu2024prompttts++} control the style of speech through a text that describes the characteristics (including gender, pitch, speaking speed, etc.). 
Since these methods typically use limited combinations of given aspects, the diversity of the synthesized speech is relatively low and it is also difficult to achieve finer control.

\begin{figure}[!t]
\centerline{\includegraphics[scale=0.5]{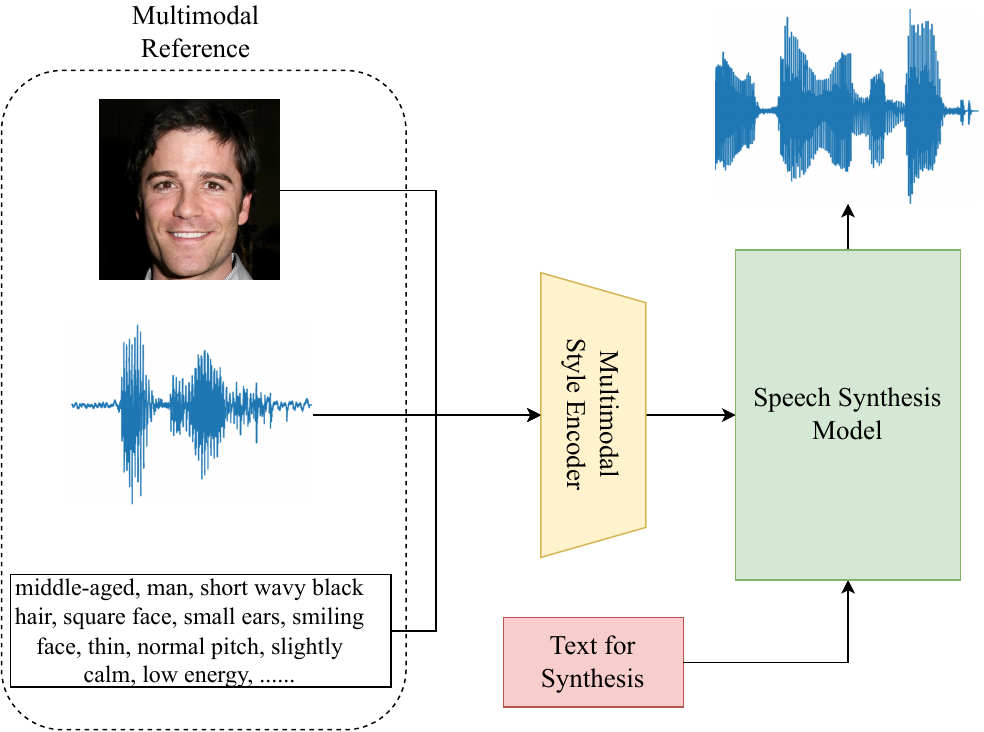}}
\caption{A unified multimodal controllable speech synthesis framework.}
\label{fig1}
\end{figure}

Since real-world speech synthesis applications often encounter scenarios where input from certain modalities may not be available, the utilization of multimodal input offers a more practical solution. 
By integrating various modalities of input, multimodal based systems can flexibly generate more appropriate speech for specific tasks. 
Fig.~\ref{fig1} illustrates the framework of a multimodal based synthesis system, where the reference input can be of any modality. 
% The multimodal encoder extracts style information from these inputs, and the speech synthesis model utilizes this extracted information to control the synthesized speech.
% Recently, both \cite{li2024mm} and \cite{guan2024mm} have introduced multimodal speech synthesis models that focus on emotional and stylistic information, respectively. 
% Nevertheless, these methods have limitations. 
% A significant drawback is their reliance on fully matched multimodal training data (speech-face-text prompt triplets), which significantly limits the quality and quantity of training data.
% This constrains the performance and applicability of the models.
Recent multimodal based speech synthesis models \cite{li2024mm,guan2024mm} focus on emotional and stylistic information.
A significant drawback is their reliance on fully matched speech-face-text prompt triplets for training. 
This dependency restricts the quality and quantity of the training data, thus limiting the performance and applicability of the model.

In this paper, we focus on speaker related style and propose a novel method to address the aforementioned issues in multimodal, face based, and text prompt based methods.
Specifically, we convert the requirement for speech-face-text prompt triplet training data into the requirements for three types of paired data and divide the training process into three stages.
During the first training stage, we align facial and speech features and enhances facial features' speaker consistency and generalization via supervised learning and knowledge distillation.
% During the first training stage, we achieve alignment between facial image features and speech features.
% With supervised learning and knowledge distillation, we enhance the speaker consistency and generalization ability of facial features. 
% And we use pretrained face recognition models and speaker verification models to perform knowledge distillation on the image encoder, in order to address the issue of generalization.
% In the second phase, we train the text encoder using both text prompt-face and text prompt-speech data simultaneously. 
% With this method, we greatly expand the available text data, no longer limited to combinations of a few characteristics, thereby enhancing the diversity of the generated speech.
In the second phase, we train the text encoder with text prompt-face and text prompt-speech data, expanding text data and increasing speech diversity.
In the third phase, we train the speech synthesis model using only speech as the reference input. 
Benefiting from the high-quality speech data, the quality of the synthesized speech has been improved.
Experimental results show that our method outperforms single-modal baseline methods in both face based and text prompt based speech synthesis.
The demo is available on the online page\footnote{\url{https://thuhcsi.github.io/icme2025-MMTTS/}}.

% In summary, the contributions of this paper are as follows:
% \begin{itemize}
%   \item We propose a multi-stage multimodal controllable speech synthesis framework to address the issue of lacking multimodal paired data and improve the quality of synthesized speech.
%   \item We use supervised learning and knowledge distillation methods to solve the consistency and generalization issues of facial features.
%   \item We train the text encoder using both text-face and text-speech data to enhance the diversity of the generated speech based on text prompts. And to the best of our knowledge, we are the first model that can generate speech with facial prompts.
% \end{itemize}

\section{Related Work}
\subsection{Face Based Speech Synthesis}
In face based speech synthesis, the facial image is used as the conditional input. 
The speaker identity characteristics expressed by the face are used to synthesize speech with similar features. 
These methods can be employed when it is not feasible to obtain speech samples from the target speaker, like when creating voices for fictional characters.

Depending on how the face information is utilized, face based speech synthesis can be broadly classified into two categories. 
(i) Two-stage methods. 
These methods generally involve two stages of training. 
In the first stage, reference information from speech or facial images is mapped into a consistent speaker embedding space. 
In the second stage, the trained speaker embeddings are used to train the speech synthesis model. 
Face2Speech\cite{goto2020face2speech} used a loss similar to GE2E\cite{wan2018generalized} loss to train the face encoder, preserving the generalization ability of the pretrained speech encoder.
% \cite{pluster2021hearing} replaced speech embedding with a global style token\cite{wang2018style}, allowing the model to learn rich style information.
In addition, Synthesees\cite{10448433} employed supervised learning to capture distinctive speaker attributes.
(ii) End-to-end methods.
These methods directly train the speech synthesis model using facial images as conditional inputs, eliminating the need for an initial alignment process.
FaceTTS\cite{10094745} incorporated additional binding loss to maintain the similarity between the generated speech and the original speech.
FVTTS\cite{lee24_interspeech} proposed an end-to-end architecture that obviates the necessity for training additional networks.
However, the synthesis performance of these methods is relatively poor due to the lack of sufficient text-speech-face triplet data.

\subsection{Text Prompt Based Speech Synthesis}
Speech synthesis based on text prompts utilizes natural language as prompts to generate expressive speech.
This approach eliminates the need for users to possess acoustic knowledge, thereby allowing them to flexibly describe the desired speech.
Recently, PromptTTS\cite{guo2023prompttts} was the first to use a text description as a prompt to guide speech generation. 
They use five distinct factors to describe speech: gender, pitch, speaking speed, volume, and emotion. 
Speech data is divided into several categories based on these characteristics and they write style prompts for each category. 
In contrast to PromptTTS, InstructTTS\cite{yang2024instructtts} does not impose restrictions on the format of style prompts. 
Users are allowed to use any natural language expressions to describe their desired speaking style.  
And to precisely control speaker identity, PromptTTS++\cite{shimizu2024prompttts++} introduced a speaker prompt that describes voice characteristics.
In this paper, we've incorporated facial descriptions into our training process, enhancing the diversity of the synthesized output.
\section{Method}
The proposed method consists of three training stages: aligning face embeddings with speech embeddings, aligning text embeddings with face and speech embeddings, and training the speech synthesis model.

\begin{figure*}[!t]
  \centering
  \subfloat[Face Encoder training process]{
    \label{fig2a}
    \includegraphics[width=0.5\textwidth]{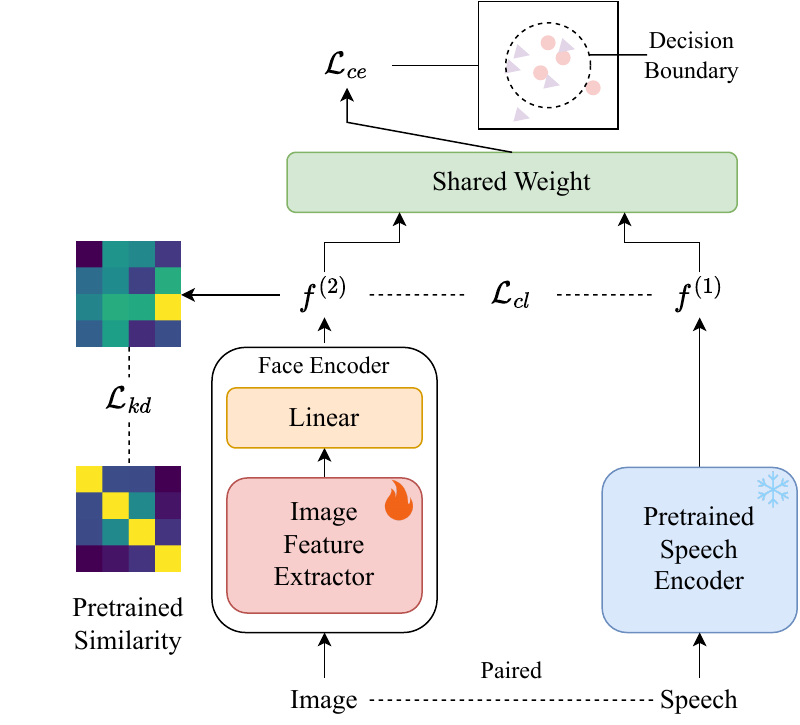}
  }
  \subfloat[Text Encoder training process]{
    \label{fig2b}
    \includegraphics[width=0.465\textwidth]{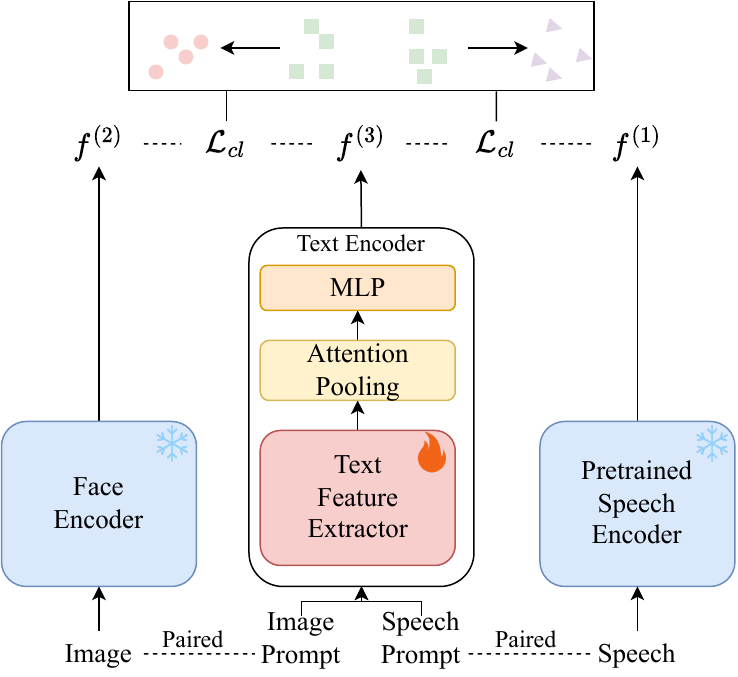}
  }
  \caption{Overview of the training process for the Face Encoder (left) and the Text Encoder (right).}
  \label{fig2}
\end{figure*}

\begin{figure*}[!t]
\centerline{\includegraphics[scale=0.6]{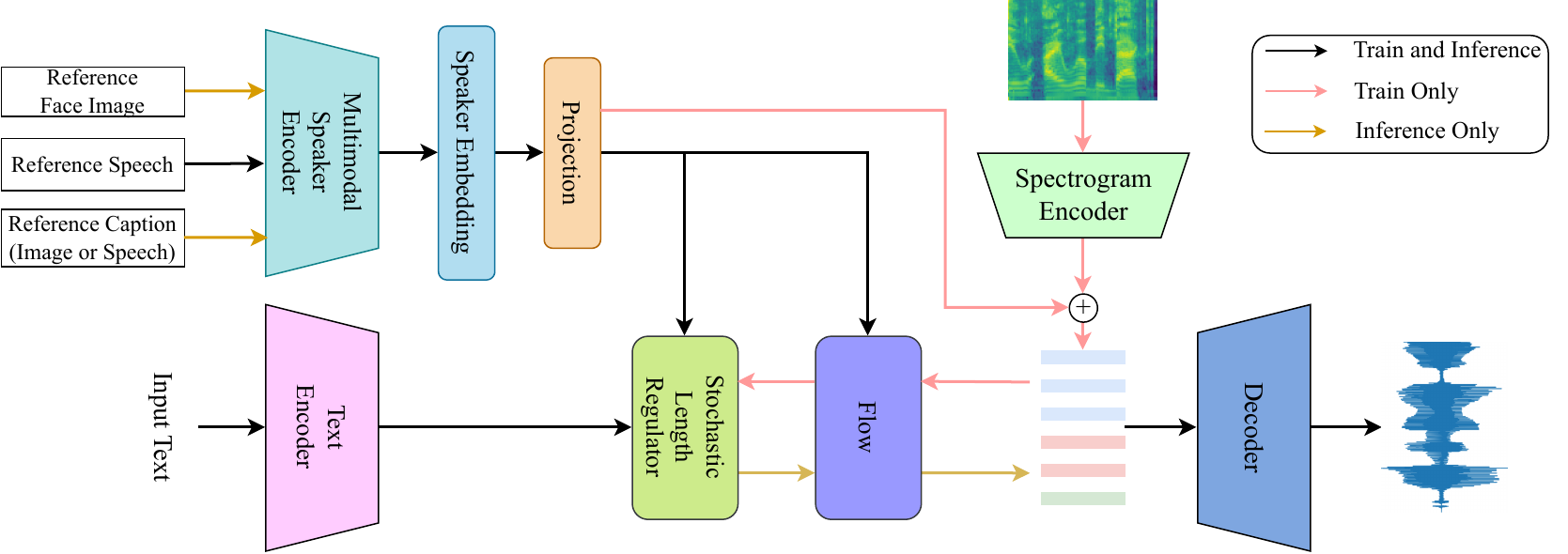}}
\caption{Training process and inference process for VITS based speech synthesis model.}
\label{fig3}
\end{figure*}

Fig.~\ref{fig2a} illustrates the training process of the face encoder, where the goal is to make the proposed face encoder share the speaker space with a pretrained, frozen speech encoder. 
By training the model using supervised learning and knowledge distillation methods, the face-based speaker representation becomes more discriminative and generalizable. 
Fig.~\ref{fig2b} shows the training process of the text encoder. 
By using both face-text prompt pairs and speech-text prompt pairs, the text encoder gains the ability to generate diverse speaker embeddings from various prompts. 
Finally, Fig.~\ref{fig3} demonstrates the training and inference process of the speech synthesis model. 
We train the speech synthesis model using high-quality speech data and the speech speaker encoder. 
During the inference stage, the proposed model can synthesize speech using reference inputs from all modalities. 
In the following sections, we will provide a detailed introduction to each training stage.
\subsection{Face-Speech Alignment}
In situations where paired data is insufficient, forcing alignment between two different modalities often faces issues with generalization. 
To train the face encoder to align with the pretrained speech encoder while improving its generalization ability, we use three different loss terms ($\mathcal{L}_{ce}$, $\mathcal{L}_{kd}$, $\mathcal{L}_{cl}$).
\subsubsection{Weight Shared Classification}
To enhance the model's robustness to different angles and lighting changes of the same speaker, we employed a classification loss during training. 
This approach leverages supervisory information from the dataset to improve the model's generalization capabilities.
However, training the face embedding directly with cross entropy loss may cause the model to rely on features from the face modality rather than features shared between modalities. 
Therefore, based on AMSoftmax\cite{wang2018additive}, we proposed a classification loss with a shared weight between modalities, which is formulated as follows:
\begin{equation}
\mathcal{L}_{ce}^{(o)}=-\frac{1}{n}\sum_{i=1}^{n}\log\frac{e^{s\cdot(W^T_{y_i}f^{(o)}_i-m)}}{e^{s\cdot(W^T_{y_i}f^{(o)}_i-m)}+\sum_{j=1,j\neq y_i}^ce^{s\cdot(W^T_{j}f^{(o)}_i)}}
\end{equation}
\begin{equation}
\mathcal{L}_{ce}=\alpha\mathcal{L}_{ce}^{(1)}+\mathcal{L}_{ce}^{(2)}
\end{equation}
where $n$ is the number of samples in mini-batch, $W\in\mathbb{R}^{d\times c}$ is a shared learnable parameter ($d$ means speaker embedding dim), $f_i^{(o)}$ is the speaker embedding in modality $o$ ($1$ for speech, $2$ for face image, $3$ for text prompt), $y_i$ is corresponding speaker label, $c$ is the number of speakers, $\alpha$ is a balance hyperparameter and $m$ and $s$ are hyperparameters to control the class margin and scale.

By using the shared weight $W$, The decision boundary is drawn to the region where face features overlap with speech features.
This allows speaker identification information derived from the speech modality to be passed to the face encoder, forcing it to learn shared speaker features.
\subsubsection{Joint Knowledge Distillation}
Pretrained face recognition models demonstrate high levels of generalization capability.
However, the representation space after modality alignment may encounter the issue of mode collapse\cite{bangalath2022bridging} (i.e., mapping different speakers to similar representations).

Knowledge distillation extracts knowledge from the teacher model and integrates it with the student model, thereby enhancing the student's knowledge and performance.
To address the issue of mode collapse, we attempt to transfer refined knowledge from pretrained models of two modalities to the face encoder. 
Given the differences in training objectives between student and teacher models, as well as among teacher models of various modalities, we construct task-independent transferable knowledge items from the teachers.

Inspired by \cite{zhang2021joint}, we utilize teacher knowledge to construct a similarity matrix, which effectively captures high-order comparisons between instances and provides richer supervision from teachers. 
Specifically, the similarity matrix is
\begin{equation}
S^{(o)}_{ij}=\cos(\hat{f}^{(o)}_i,\hat{f}^{(o)}_j)
\end{equation}
where $\hat{f}^{(o)}_i$ is the embedding from the pretrained teacher network of modality $o$, $\cos(\cdot ,\cdot)$ is the cosine similarity.
Similarity matrices from different modalities are further merged into a unified similarity matrix:
\begin{equation}
S_{ij}=\mu S^{(1)} + (1-\mu) S^{(2)}, \mu\in (0,1)
\end{equation}
where$\mu$ is a balance parameter.
With the similarity matrix, we have the knowledge distillation loss as:
\begin{equation}
\begin{aligned}
\mathcal{L}_{kd}&=\sum_{i=1}^n\sum_{j=1}^n(\vert\vert S_{ij}-\cos(f_i^{(1)},f_j^{(2)})\vert\vert
\\&+\beta\vert\vert S_{ij}-\cos(f_i^{(2)},f_j^{(2)})\vert\vert)
\end{aligned}  
\end{equation}
where $\beta$ is a hyperparameter.

The knowledge distillation loss term ensures that the model retains the similarity between different modalities as well as within the same modality from the teacher model during the training process, which improves the model's generalization ability.
\subsubsection{Contrastive Learning Alignment}
Contrastive learning\cite{oord2018representation} is a prevalent approach in multimodal alignment. 
It aligns embeddings from different modalities within a unified space by encouraging similarity between embeddings from the same speaker while ensuring differences between embeddings from different speakers. 
Formally, 
\begin{equation}
\mathcal{L}_{cl}=\sum_{i=1}^n-\log\frac{e^{\cos(f_i^{(1)},f_{i}^{(2)})/\tau}}{e^{\cos(f_i^{(1)},f_{i}^{(2)})/\tau}+\sum_{f_{i^-}^{(2)}\in\mathcal{N}_i}e^{\cos(f_i^{(1)},f_{i^-}^{(2)})/\tau}}
\end{equation}
where $\mathcal{N}_i$ is the set of negative samples corresponding to $f_i^{(2)}$, $\tau$ is a temperature parameter.
Through contrastive learning, the model can learn more universal representations, enabling it to adapt to new data and improving its flexibility and practicality.

Finally, the training loss of the face encoder is:
\begin{equation}
\mathcal{L}=\mathcal{L}_{ce}+\gamma\mathcal{L}_{kd}+\mathcal{L}_{cl}
\end{equation}
where $\gamma$ is a hyperparameter to balance loss terms.
\subsection{Text Prompt Alignment}
Once the speech and face encoders have been obtained, we can employ them to train the text encoder.
The architecture of text encoder is shown in Fig.~\ref{fig2b}.
We use a transformer encoder model as the backbone of text feature extractor.
After that, we utilize an attention pooling layer followed by multilayer perceptron (MLP) to obtain a sentence level embedding.

During the training phase, data from three modalities (text prompt, face, and speech) are utilized simultaneously.
Specifically, we first obtain face-text prompt pairs and speech-text prompt pairs. 
Then, we align the text embeddings with the face embeddings and speech embeddings, respectively, using the previously described alignment loss.
\begin{equation}
\mathcal{L}_{cl}=\sum_{i=1}^{2n}-\log\frac{e^{\cos(f_i^{(3)},f_{i}^{(*)})/\tau}}{e^{\cos(f_i^{(3)},f_{i}^{(*)})/\tau}+\sum_{f_{i^-}^{(*)}\in\mathcal{N}_i}e^{\cos(f_i^{(3)},f_{i^-}^{(*)})/\tau}}
\end{equation}
where $f_{i}^{(*)}$ represents the corresponding embedding of the speech or face modality.

\subsection{Speech Synthesis}
The proposed architecture is compatible with various speech synthesis models. 
In this paper, we use the end-to-end VITS\cite{kim2021conditional} model.
VITS incorporates normalizing flows and employs adversarial training processes, which enhances its expressiveness in generative speech modeling. 
Additionally, VITS introduces a stochastic duration predictor to synthesize speech with varying rhythms from the input text.

During the training phase, we only use the speech speaker encoder to ensure model performance. 
In the inference phase, we can utilize aligned encoders from various modalities to achieve personalized speech synthesis.

\section{Experiments}
\subsection{Experiments Settings}
For face-speech alignment training, we use the widely used LRS3 dataset\cite{afouras2018lrs3}. 
It contains a training set with 4,004 speakers and a test set with 457 speakers.
For text encoder training, we utilized two datasets: the face caption dataset CelebAText-HQ\cite{10.1145/3474085.3475391} and the speech caption dataset LibriTTS-P\cite{kawamura2024libritts}.
CelebAText-HQ consists of 15,010 facial images, each containing 10 manually annotated captions describing the face.
For simplicity, we employ a large language model to extract keywords from each caption sentence and combine the keywords into a sequence as the prompt.
LibriTTS-P is a dataset that is based on LibriTTS-R\cite{koizumi2023libritts}, which includes utterance-level prompts of speaking style and speaker-level prompts characterizing the speaker's features. 
And it comprises 2,443 speakers and a total of 373,868 captions.
The speech synthesis model was trained using LibriTTS\cite{zen2019libritts}.

In the first phase of training, we initialize the image feature extractor with Facenet\cite{schroff2015facenet}, which is pretrained on VGGFace2\cite{cao2018vggface2}.
And a multi-task cascaded neural network\cite{zhang2016joint} (MTCNN) is utilized for face detection and alignment.
To make the model more robust, we performed data augmentations on the input facial images, including random rotation and random color jitter. 
For speech encoder, we use an ECAPA-TDNN\cite{desplanques2020ecapa} model pretrained with the VoxCeleb2\cite{chung2018voxceleb2}.
In the optimization process, we use the Adam optimizer with a batch size of 96 and a learning rate of 0.0002 for a total step of 200,000.
The hyperparameters are set to the following values: $\alpha=0.1$, $m=0.2$, $s=30$, $\mu=0.8$, $\beta=0.1$, $\tau=0.1$, $\gamma=10$.

In the second phase of training, a pretrained T5\cite{raffel2020exploring} encoder serves as the initial text feature extractor.
For each step, a batch consisting of 64 speech-text prompt pairs and 64 face-text prompt pairs is fed into the model.
Similar to the first phase, we employ the Adam optimizer with a learning rate of 0.0002 for a total of 200,000 steps.
For speech synthesis model, we trained 600,000 steps with a batch size of 96 and a learning rate of 0.0002.

To verify the effectiveness of proposed method, We compared it with several baseline models. 
For face based speech synthesis, we use FaceTTS, Synthesees and Face2Speech.
The provided pretrained weights were utilized for FaceTTS.
Synthesees and Face2Speech were re-implemented and the speech synthesis model was replaced with VITS for a fair comparison.
Since the training goal of the face encoder is to align with the speech encoder, we also use the reference speech based method as an upper bound for comparison.
For text prompt based speech synthesis, we use pretrained PromptTTS++ as baseline model, which also adopts style prompt and speaker prompt.
\subsection{Evaluation Metrics}
We have conducted substantial experiments to verify the performance of speech based speech synthesis.
In the objective evaluation, the equal error rate (EER) and minimum detection cost function (minDCF), which are commonly employed in speaker verification, were utilized to assess the discriminative ability of speaker representations.
And to verify the naturalness of the model-generated audio and its similarity to the reference, we conducted subjective evaluations: 5-scale naturalness Mean Opinion Score (MOS) and speaker consistency MOS tests.

For text prompt based speech synthesis, in addition to subjective metrics, we also use silhouette score.
The speaker embeddings of the generated samples were divided into two clusters based on gender. 
Through the silhouette score, we quantify the cohesion and distinction within the data clusters.
A lower contour coefficient indicates a larger distribution area for the cluster, thus demonstrating superior diversity.
\begin{table}[!ht]
    \centering
    \caption{Objective results for face based speech synthesis}
    \resizebox{0.7\linewidth}{!}{
    \begin{tabular}{ccc}
    \hline
        Methods & EER (\%) $\downarrow$ & MinDCF $\downarrow$ \\ \hline
        Reference Speech & 2.5445 & 0.1559 \\ \hline
        Face2Speech & 11.0522 & 0.4594 \\ 
        Synthesees & 8.2273 & 0.4684 \\ 
        Ours & \textbf{4.5801} & \textbf{0.2797} \\ \hline
    \end{tabular}}
\label{tab:1}
\end{table}
\subsection{Results of Face Based Speech Synthesis}
\subsubsection{Objective Results}
Tab.~\ref{tab:1} presents the EER and MinDCF results for speaker representations derived from facial images and speech.
The results show that our method significantly outperforms the baseline method in terms of discriminability and generalization for unseen speakers.
The Face2Speech method, which only uses an alignment loss, achieved the worst results, indicating that utilizing label information from facial datasets for supervision is beneficial.
Moreover, our method has also shown improvement over the Synthesees method, which uses label information as well. 
This suggests that leveraging pretrained face recognition models for knowledge distillation can enhance the model's generalization capability.
\subsubsection{Ablation Study}
Tab.~\ref{tab:2} presents the ablation experiment results for loss terms. 
Specifically, in the ablation of the contrastive learning loss, we replaced the alignment loss with a cosine similarity loss.

\begin{table}[!ht]
    \centering
    \caption{Ablation results for face based speech synthesis}
    \resizebox{0.7\linewidth}{!}{
    \begin{tabular}{ccc}
    \hline
        Methods & EER (\%) $\downarrow$ & MinDCF $\downarrow$ \\ \hline
        w/o $\mathcal{L}_{ce}$ & 9.5843 & 0.4052 \\ 
        w/o $\mathcal{L}_{kd}$ & 10.6870 & 0.4823 \\ 
        w/o $\mathcal{L}_{cl}$ & 8.0242 & 0.3864 \\ \hline
    \end{tabular}}
\label{tab:2}
\end{table}
From the results, it can be seen that replacing the contrastive learning loss has a relatively minor impact on performance, although there is still a notable decline. 
These results demonstrate the efficacy of our proposed supervised learning method and knowledge distillation method, while also highlighting the superiority of contrastive learning methods in representation learning tasks.
Removing knowledge distillation has the most significant impact on the results, yet removing supervised learning still achieves performance comparable to the baseline. 
This indicates that the model can retain more knowledge about discrimination through knowledge distillation, thereby enhancing its capacity for generalization.
\subsubsection{Subjective Results}
Tab.~\ref{tab:3}  displays the results for the naturalness (N-MOS) and similarity (S-MOS) of synthesized speech.
Our method achieved the highest naturalness, and this result proved that the three-stage training strategy with high-quality training data can produce more natural speech.  
The N-MOS of FaceTTS indicates that it is challenging to synthesize expected speech end-to-end with a small and low-quality face-speech-text dataset.
Moreover, our method has also improved in speaker similarity, which verifies the effectiveness of capturing speech related style attributes from facial images and synthesizing more appropriate speech.
\begin{table}[!ht]
    \centering
    \caption{Subjective results for face based speech synthesis}
    \resizebox{0.7\linewidth}{!}{
    \begin{tabular}{ccc}
    \hline
        Methods & N-MOS $\uparrow$ & S-MOS $\uparrow$ \\ \hline
        Ground Truth & 4.31$\pm$0.09 & 4.32$\pm$0.10 \\ 
        Reference Speech & 3.55$\pm$0.10 & 3.58$\pm$0.10 \\ \hline
        FaceTTS & 2.36$\pm$0.10 & 2.18$\pm$0.12 \\ 
        Synthesees & 2.94$\pm$0.11 & 2.69$\pm$0.13 \\ 
        Ours & \textbf{3.48$\pm$0.11} & \textbf{3.63$\pm$0.10} \\ \hline
    \end{tabular}}
\label{tab:3}
\end{table}
\subsection{Results of Text Prompt based Speech Synthesis}
Tab.~\ref{tab:4} presents the experimental results of text-based speech synthesis. 
Compared to the baseline method, our method has achieved certain improvements in both speech naturalness and matching with the text prompt. 
The findings demonstrate that our method exhibits robust capabilities in natural language modeling and the ability to capture style information.
\begin{table}[!ht]
    \centering
    \caption{Text prompt based speech synthesis results}
    \resizebox{0.9\linewidth}{!}{
    \begin{tabular}{cccc}
    \hline
        Methods & N-MOS $\uparrow$ & S-MOS $\uparrow$ & Silhouette Score $\downarrow$\\ \hline
        PromptTTS++ & 3.42$\pm$0.08 & 3.23$\pm$0.09 & 0.1365\\ 
        Ours & \textbf{3.62$\pm$0.08} & \textbf{3.75$\pm$0.09} & \textbf{0.0904}\\ \hline
    \end{tabular}}
\label{tab:4}
\end{table}
Moreover, our method offers distinct advantages in silhouette score. 
This indicates that the synthesized speaker distribution is broader, which suggests that our method is effective in generating more diverse speeches with the usage of both facial prompts and speech prompts.

\section{Conclusion}
In this paper, we introduce a multi-stage multimodal controllable speech synthesis framework to address the limitations posed by the fully matched training data and to enhance the quality of synthesized speech. 
We employs supervised learning and knowledge distillation to address inconsistencies and enhance the generalization of facial features. 
Furthermore, the text encoder is trained on a combination of text prompt-face and text prompt-speech data, thereby increasing the diversity of the generated speech. 
The experimental findings demonstrate that our method outperforms single-modal baseline models in both face based and text prompt based speech synthesis, substantiating its efficacy in generating high-quality speech.
% \section*{Acknowledgment}

% The preferred spelling of the word ``acknowledgment'' in America is without 
% an ``e'' after the ``g''. Avoid the stilted expression ``one of us (R. B. 
% G.) thanks $\ldots$''. Instead, try ``R. B. G. thanks$\ldots$''. Put sponsor 
% acknowledgments in the unnumbered footnote on the first page.

\bibliographystyle{IEEEbib}
\bibliography{icme2025references}

\begin{thebibliography}{10}

\bibitem{jiang2024mega}
Ziyue Jiang, Jinglin Liu, Yi~Ren, Jinzheng He, Zhenhui Ye, Shengpeng Ji, Qian Yang, Chen Zhang, Pengfei Wei, Chunfeng Wang, et~al.,
\newblock ``Mega-tts 2: Boosting prompting mechanisms for zero-shot speech synthesis,''
\newblock in {\em The Twelfth International Conference on Learning Representations}, 2024.

\bibitem{du2024cosyvoice}
Zhihao Du, Qian Chen, Shiliang Zhang, Kai Hu, Heng Lu, Yexin Yang, Hangrui Hu, Siqi Zheng, Yue Gu, Ziyang Ma, et~al.,
\newblock ``Cosyvoice: A scalable multilingual zero-shot text-to-speech synthesizer based on supervised semantic tokens,''
\newblock {\em arXiv preprint arXiv:2407.05407}, 2024.

\bibitem{wang2024maskgct}
Yuancheng Wang, Haoyue Zhan, Liwei Liu, Ruihong Zeng, Haotian Guo, Jiachen Zheng, Qiang Zhang, Xueyao Zhang, Shunsi Zhang, and Zhizheng Wu,
\newblock ``Maskgct: Zero-shot text-to-speech with masked generative codec transformer,''
\newblock {\em arXiv preprint arXiv:2409.00750}, 2024.

\bibitem{10094745}
Jiyoung Lee, Joon Son~Chung, and Soo-Whan Chung,
\newblock ``Imaginary voice: {{Face-styled}} diffusion model for text-to-speech,''
\newblock in {\em {{ICASSP}} 2023 - 2023 {{IEEE}} International Conference on Acoustics, Speech and Signal Processing ({{ICASSP}})}, 2023, pp. 1--5.

\bibitem{10448433}
Jae~Hyun Park, Joon-Gyu Maeng, TaeJun Bak, and Young-Sun Joo,
\newblock ``{{SYNTHE-SEES}}: {{Face}} based text-to-speech for virtual speaker,''
\newblock in {\em {{ICASSP}} 2024 - 2024 {{IEEE}} International Conference on Acoustics, Speech and Signal Processing ({{ICASSP}})}, 2024, pp. 10321--10325.

\bibitem{goto2020face2speech}
Shunsuke Goto, Kotaro Onishi, Yuki Saito, Kentaro Tachibana, and Koichiro Mori,
\newblock ``{{Face2Speech}}: {{Towards}} multi-speaker text-to-speech synthesis using an embedding vector predicted from a face image.,''
\newblock in {\em {{INTERSPEECH}}}, 2020, pp. 1321--1325.

\bibitem{lee24_interspeech}
Minyoung Lee, Eunil Park, and Sungeun Hong,
\newblock ``{{FVTTS}} : {{Face}} based voice synthesis for text-to-speech,''
\newblock in {\em Interspeech 2024}, 2024, pp. 4953--4957.

\bibitem{yang2024instructtts}
Dongchao Yang, Songxiang Liu, Rongjie Huang, Chao Weng, and Helen Meng,
\newblock ``Instructtts: {{Modelling}} expressive tts in discrete latent space with natural language style prompt,''
\newblock {\em IEEE/ACM Transactions on Audio, Speech, and Language Processing}, 2024.

\bibitem{guo2023prompttts}
Zhifang Guo, Yichong Leng, Yihan Wu, Sheng Zhao, and Xu~Tan,
\newblock ``Prompttts: {{Controllable}} text-to-speech with text descriptions,''
\newblock in {\em {{ICASSP}} 2023-2023 {{IEEE}} International Conference on Acoustics, Speech and Signal Processing ({{ICASSP}})}. 2023, pp. 1--5, IEEE.

\bibitem{shimizu2024prompttts++}
Reo Shimizu, Ryuichi Yamamoto, Masaya Kawamura, Yuma Shirahata, Hironori Doi, Tatsuya Komatsu, and Kentaro Tachibana,
\newblock ``{{PromptTTS}}++: {{Controlling}} speaker identity in prompt-based text-to-speech using natural language descriptions,''
\newblock in {\em {{ICASSP}} 2024-2024 {{IEEE}} International Conference on Acoustics, Speech and Signal Processing ({{ICASSP}})}. 2024, pp. 12672--12676, IEEE.

\bibitem{li2024mm}
Xiang Li, Zhi-Qi Cheng, Jun-Yan He, Xiaojiang Peng, and Alexander~G Hauptmann,
\newblock ``Mm-tts: {{A}} unified framework for multimodal, prompt-induced emotional text-to-speech synthesis,''
\newblock {\em arXiv preprint arXiv:2404.18398}, 2024.

\bibitem{guan2024mm}
Wenhao Guan, Yishuang Li, Tao Li, Hukai Huang, Feng Wang, Jiayan Lin, Lingyan Huang, Lin Li, and Qingyang Hong,
\newblock ``{{MM-TTS}}: {{Multi-modal}} prompt based style transfer for expressive text-to-speech synthesis,''
\newblock in {\em Proceedings of the {{AAAI}} Conference on Artificial Intelligence}, 2024, vol.~38, pp. 18117--18125.

\bibitem{wan2018generalized}
Li~Wan, Quan Wang, Alan Papir, and Ignacio~Lopez Moreno,
\newblock ``Generalized end-to-end loss for speaker verification,''
\newblock in {\em 2018 IEEE International Conference on Acoustics, Speech and Signal Processing (ICASSP)}. IEEE, 2018, pp. 4879--4883.

\bibitem{wang2018additive}
Feng Wang, Jian Cheng, Weiyang Liu, and Haijun Liu,
\newblock ``Additive margin softmax for face verification,''
\newblock {\em IEEE Signal Processing Letters}, vol. 25, no. 7, pp. 926--930, 2018.

\bibitem{bangalath2022bridging}
Hanoona Bangalath, Muhammad Maaz, Muhammad~Uzair Khattak, Salman~H Khan, and Fahad Shahbaz~Khan,
\newblock ``Bridging the gap between object and image-level representations for open-vocabulary detection,''
\newblock {\em Advances in Neural Information Processing Systems}, vol. 35, pp. 33781--33794, 2022.

\bibitem{zhang2021joint}
Peng-Fei Zhang, Jiasheng Duan, Zi~Huang, and Hongzhi Yin,
\newblock ``Joint-teaching: {{Learning}} to refine knowledge for resource-constrained unsupervised cross-modal retrieval,''
\newblock in {\em Proceedings of the 29th {{ACM}} International Conference on Multimedia}, 2021, pp. 1517--1525.

\bibitem{oord2018representation}
Aaron van~den Oord, Yazhe Li, and Oriol Vinyals,
\newblock ``Representation learning with contrastive predictive coding,''
\newblock {\em arXiv preprint arXiv:1807.03748}, 2018.

\bibitem{kim2021conditional}
Jaehyeon Kim, Jungil Kong, and Juhee Son,
\newblock ``Conditional variational autoencoder with adversarial learning for end-to-end text-to-speech,''
\newblock in {\em International Conference on Machine Learning}. 2021, pp. 5530--5540, PMLR.

\bibitem{afouras2018lrs3}
Triantafyllos Afouras, Joon~Son Chung, and Andrew Zisserman,
\newblock ``Lrs3-ted: a large-scale dataset for visual speech recognition,''
\newblock {\em arXiv preprint arXiv:1809.00496}, 2018.

\bibitem{10.1145/3474085.3475391}
Jianxin Sun, Qi~Li, Weining Wang, Jian Zhao, and Zhenan Sun,
\newblock ``Multi-caption text-to-face synthesis: {{Dataset}} and algorithm,''
\newblock in {\em Proceedings of the 29th {{ACM}} International Conference on Multimedia}, New York, NY, USA, 2021, Mm '21, pp. 2290--2298, Association for Computing Machinery.

\bibitem{kawamura2024libritts}
Masaya Kawamura, Ryuichi Yamamoto, Yuma Shirahata, Takuya Hasumi, and Kentaro Tachibana,
\newblock ``{{LibriTTS-p}}: A corpus with speaking style and speaker identity prompts for text-to-speech and style captioning,''
\newblock {\em arXiv preprint arXiv:2406.07969}, 2024.

\bibitem{koizumi2023libritts}
Yuma Koizumi, Heiga Zen, Shigeki Karita, Yifan Ding, Kohei Yatabe, Nobuyuki Morioka, Michiel Bacchiani, Yu~Zhang, Wei Han, and Ankur Bapna,
\newblock ``Libritts-r: A restored multi-speaker text-to-speech corpus,''
\newblock {\em arXiv preprint arXiv:2305.18802}, 2023.

\bibitem{zen2019libritts}
Heiga Zen, Viet Dang, Rob Clark, Yu~Zhang, Ron~J Weiss, Ye~Jia, Zhifeng Chen, and Yonghui Wu,
\newblock ``Libritts: A corpus derived from librispeech for text-to-speech,''
\newblock {\em arXiv preprint arXiv:1904.02882}, 2019.

\bibitem{schroff2015facenet}
Florian Schroff, Dmitry Kalenichenko, and James Philbin,
\newblock ``Facenet: A unified embedding for face recognition and clustering,''
\newblock in {\em Proceedings of the IEEE conference on computer vision and pattern recognition}, 2015, pp. 815--823.

\bibitem{cao2018vggface2}
Qiong Cao, Li~Shen, Weidi Xie, Omkar~M Parkhi, and Andrew Zisserman,
\newblock ``Vggface2: A dataset for recognising faces across pose and age,''
\newblock in {\em 2018 13th IEEE international conference on automatic face \& gesture recognition (FG 2018)}. IEEE, 2018, pp. 67--74.

\bibitem{zhang2016joint}
Kaipeng Zhang, Zhanpeng Zhang, Zhifeng Li, and Yu~Qiao,
\newblock ``Joint face detection and alignment using multitask cascaded convolutional networks,''
\newblock {\em IEEE signal processing letters}, vol. 23, no. 10, pp. 1499--1503, 2016.

\bibitem{desplanques2020ecapa}
Brecht Desplanques, Jenthe Thienpondt, and Kris Demuynck,
\newblock ``Ecapa-tdnn: Emphasized channel attention, propagation and aggregation in tdnn based speaker verification,''
\newblock {\em arXiv preprint arXiv:2005.07143}, 2020.

\bibitem{chung2018voxceleb2}
Joon~Son Chung, Arsha Nagrani, and Andrew Zisserman,
\newblock ``Voxceleb2: Deep speaker recognition,''
\newblock {\em arXiv preprint arXiv:1806.05622}, 2018.

\bibitem{raffel2020exploring}
Colin Raffel, Noam Shazeer, Adam Roberts, Katherine Lee, Sharan Narang, Michael Matena, Yanqi Zhou, Wei Li, and Peter~J Liu,
\newblock ``Exploring the limits of transfer learning with a unified text-to-text transformer,''
\newblock {\em Journal of machine learning research}, vol. 21, no. 140, pp. 1--67, 2020.

\end{thebibliography}

\end{document}